\documentclass[review,12pt,sort&compress]{elsarticle}

\usepackage{amssymb}

\usepackage{color} 
\usepackage{epstopdf}
\usepackage{float} 
\usepackage{subfig}
\usepackage{booktabs}
\usepackage{tabularx} 
\usepackage{subfiles} 
\usepackage{multirow}
\usepackage{hyperref}
\usepackage{enumitem}
\usepackage{diagbox}
\usepackage{makecell}

\journal{Powder Technology}

\begin{document}

\begin{frontmatter}



\title{Resolved CFD-DEM simulations of the hydraulic conveying of coarse grains through a very-narrow elbow. \tnoteref{label_note_copyright} \tnoteref{label_note_doi}}

\tnotetext[label_note_copyright]{\copyright 2019. This manuscript version is made available under the CC-BY-NC-ND 4.0 license http://creativecommons.org/licenses/by-nc-nd/4.0/}

\tnotetext[label_note_doi]{Accepted Manuscript for Powder Technology, v. 395, p. 811-821, 2022, DOI:10.1016/j.powtec.2021.10.022}




\author{Elmar Anton Schnorr Filho}
\author{Nicolao Cerqueira Lima}
\author[label_erick]{Erick Franklin}
\address{School of Mechanical Engineering, University of Campinas - UNICAMP\\
Rua Mendeleyev, 200, Campinas - SP, CEP: 13083-860\\
Brazil}
\address[label_erick]{phone: +55 19 35213375\\
e-mail: erick.franklin@unicamp.br, Corresponding Author}

\begin{abstract}
This paper investigates numerically the hydraulic conveying of solids through a 90$^{\circ}$ elbow that changes the flow direction from horizontal to vertical, in the very-narrow case where the ratio of pipe to particle diameters is less than 5. We performed resolved CFD-DEM (computational fluid dynamics - discrete element method) computations, in which we made use of the IB (immersed boundary) method of the open-source code CFDEM. We investigate the effects of the water flow and particle injection rate on the transport rate and sedimentation by tracking the granular structures appearing in the pipe, the motion of individual particles, and the contact network of settled particles. We found the saturated transport rate for each water velocity and that a large number of particles settle in the elbow region for smaller velocities, forming a crystal-like lattice that persists in time, and we propose a procedure to mitigate the problem.
\end{abstract}


\end{frontmatter}


\section{Introduction}
\label{sec:Introduction}

The hydraulic conveying of solid particles has been used for decades in industry as an effective way to extract and/or transport solids continuously within different facilities, being employed, for example, in mining, oil, chemical and food industries, and in wastewater treatment. Basically, it consists of pumping a liquid, usually water, that entrains solid particles as it flows through a tube or channel \cite{uzi2018flow}. It is a viable and efficient alternative to trucks, rails and belts for transporting continuously high amounts of commodities, such as coal, bitumen, ores and other grains, and also for conveying organic matter to and from bioreactors and through sewer systems. Because these materials appear in a broad range of sizes, shapes and physical properties, it is not uncommon to find particles with sizes comparable to that of pipes, with even the occurrence of blockages and clogging \cite{Cunez3, Cunez4}. In the specific case of organic matter, particles usually grow along time and develop an organic film with adhesion properties \cite{Dempsey}, increasing the probabilities of clustering, and thus blockages. In addition, piping systems frequently contain horizontal and vertical portions and changes of direction via elbows and tees, making the design of conveying systems complex. Therefore, the prediction of flow patterns, pressure drops, sedimentation and erosion rates, transport rates, clustering, clogging, and jamming in liquid-solid piping systems remains challenging \cite{vaezi2018application, Zhou4}.

In addition to pattern formation and instabilities, curved pipes can enhance particle-wall and particle-particle collisions, leading to pipe erosion and degradation of the product being transported \cite{uzi2018flow,parsi2014comprehensive,li2019relationship}. Therefore, a great part of previous studies on pneumatic and hydraulic conveying through curved pipes investigated pipe erosion and pressure drop. For example, Bourgoyne Jr \cite{bourgoyne1989experimental} investigated experimentally how sand particles being conveyed by gas and liquid in diverter systems erode different geometries of tees and bends, finding that erosion rates are two orders of magnitude higher when the fluid is a gas. Later, Shirazi et al. \cite{shirazi1995procedure} proposed an estimation model to predict wear in tees and elbows for small particle concentrations (2-3\% in weight) that agreed well with experiments. Using CFD-DEM (computational fluid dynamics - discrete element method), Zhang et al. \cite{zhang2012numerical} investigated the erosion of a slurry flow in a 90$^{\circ}$ elbow by changing the flow velocity and elbow orientation, and found that the force (impulse) related with the impact of the particle upon the wall depends on the fluid velocity, while the puncture position depends on the elbow orientation. Similarly, Peng and Cao \cite{peng2016numerical} studied numerically the erosion in a 90$^{\circ}$ elbow, but for a gas-solid flow, and found that the profile of the particle concentration is directed linked to the erosion profile.

Other aspect influencing the hydraulic conveying of grains is the ratio between the pipe and particle diameters ($D$ and $d$, respectively). When under high confinement, such as in narrow (10 $\lesssim$ $D/d$ $\lesssim$ 100) and very-narrow tubes ($D/d$ $\lesssim$ 10), different granular structures are observed \cite{Cunez, Cunez2, Cunez3}. Ravelet et al. \cite{Ravelet} investigated experimentally the hydraulic conveying of coarse grains in horizontal pipes for 6.3 $\leq$ $D/d$ $\leq$ 20 and 2.50 $\leq$ $S$ $\leq$ 3.65, where $S$ = $\rho_s$/$\rho$, $\rho_s$ being the density of grains and $\rho$ that of liquid. The authors characterized flow regimes based on visualization and pressure drop measurements, and found that the hydraulic gradient decreases with the particle size, different from commonly reported for vertical conveying. Vlasak et al. \cite{Vlasak} investigated experimentally the conveying of coarse particles in inclined tubes for $D/d$ $\approx$ 8.7 and $S$ = 2.90 by using high-speed movies and gamma-ray density meters, and found that particles are stratified and move basically as bedload \cite{Bagnold_4}. Later, Zhou et al. \cite{Zhou5} made use of unresolved CFD-DEM computations to investigate numerically the flow regimes and corresponding pressure gradients in the hydraulic conveying of coarse grains in horizontal pipes for $S$ = 2.65 and $D/d$ = 21.8 and 14.9. They found that as the water velocity is increased, flow regimes vary from stationary bed to bedload, and from the latter to heterogeneous suspension, and they proposed a phase diagram for flow regimes as functions of forces on grains. Using similar computations, Zhou et al. \cite{Zhou4} investigated numerically the hydraulic conveying of coarse grains through vertical pipes for $S$ = 2.45 and 8.0 $\leq$ $D/d$ $\leq$ 13.2. Among the obtained results, they showed that pressure drop and dispersion of particle distribution increase with both the insertion rate of particles and fluid velocity, and that the flow regime and pressure drop do not depend on the particle diameter. The latter result is different from expected, and remains to be verified.

Although appearing in flows with coarse grains in narrow and very narrow tubes, the granular structures were only fully scrutinized in a series of papers on fluidized beds. C\'u\~nez and Franklin \cite{Cunez, Cunez2, Cunez3} investigated experimentally and numerically some of those features appearing in solid-liquid fluidized beds (SLFBs) with 3.2 $\leq$ $D/d$ $\leq$ 5.3 (very-narrow case) and 2.50 $\leq$ $S$ $\leq$ 3.69. The experiments made use of high-speed movies and the numerical simulations consisted of unresolved CFD-DEM computations. The authors showed the appearance of granular plugs that present dense networks of contact forces that percolate within all the bed cross section and reach the tube wall, being thus a consequence of the high confining environment promoted by the very-narrow case. The authors also showed that very-narrow SLFBs under partial de-fluidization and re-fluidization are prone to: (i) crystallization, when a static (at the macroscopic scale) lattice with small oscillations of individual particles (microscopic motion) is formed; and (ii) jamming, when even the microscopic motion virtually stops. Their results showed the occurrence of crystallization and jamming for fluid velocities higher than those for minimum fluidization, supposed to maintain the bed fluidized (something showed previously by Goldman and Swinney \cite{Goldman} for narrow SLFBs). Later, C\'u\~nez et al. \cite{Cunez4} investigated very-narrow SLFBs consisting of bonded spheres (duos and trios) where each sphere had $D/d$ = 5.3, so that confinement was higher than in previous cases. Among other results, they showed that jamming occurs suddenly for trios in upper regions in the tube and for water velocities well above those for fluidization. For the hydraulic conveying of coarse grains, the appearance of any of those features imply serious problems. However, none of the previous studies on hydraulic conveying investigated them.

The transport of grains through elbows and bends, although commonly found in industry, is far from being completely understood, mainly under high confinement. This paper presents a numerical investigation of the hydraulic conveying of coarse grains through a pipe curved by a 90$^{\circ}$ elbow in the very narrow case, and we focus on organic particles (density close to that of water). The simulated system consisted of a 25.4-mm-ID pipe, with a 2.5$D$-long horizontal section, followed by a 90$^{\circ}$ elbow with a radius of curvature of 1D, and a 5$D$-long vertical section. Particles with $d$ = 6 mm and $\rho_s$ = $1140$ kg/$m^3$ ($D/d$ = 4.23 and $S$ = 1.14) were randomly generated at the inlet and transported by an imposed flow of water. We performed resolved CFD-DEM computations making use of the IB (immersed boundary) method of the open-source code CFDEM (www.cfdem.com) \cite{Goniva}, which couples the open-source codes OpenFOAM (www.openfoam.com) and LIGGGHTS \cite{Kloss, Berger} for the CFD and DEM computations, respectively. We investigate the effects of the water flow and particle injection rate on the transport rates and sedimentation of grains. For that, we vary the water velocities within 0.06 and and 0.10 m/s and the particle insertion rate between one-quarter-saturated and fully-saturated conditions, and identify and track the granular structures appearing in the pipe, the motion of individual particles, and the contact network of settled particles. In addition, we show how the mean water flow varies along the elbow. We find the saturated transport rate for each water velocity, and show that the mean particle velocity is an increasing function of the particle insertion rate, while particle settling in the elbow region decreases with the insertion rate. We also find that a large number of particles settles in the elbow region for smaller fluid velocities, forming a crystal-like lattice that persists in time, and we propose a simple procedure to mitigate the problem. These results can be used to avoid settling and grain accumulation in elbows and to improve the hydraulic conveying of coarse grains in industrial facilities.

The next sections present the governing equations, numerical setup, results, and the conclusions.

\section{Governing equations}
\label{sec:equations}

Our numerical simulations are of the Lagrangian-Eulerian type, where the motion of grains is computed in a Lagrangian framework and that of fluid in an Eulerian framework. We perform CFD-DEM computations, whose main equations are described next.

The solid particles are followed in a Lagrangian way by solving, for each particle, the linear and angular momentum equations, given by Eqs. \ref{Fp} and \ref{Tp}, respectively,

\begin{equation}
	m_{i}\frac{d\vec{u_{i}}}{dt}  = \vec{F}_{f,i} + m_{i}\vec{g} +
	\sum_{j \neq i}^{Np} \left( \vec{F}_{c,ij} \right) +
	\sum_{k \neq i}^{Nw} \left( \vec{F}_{c,ik} \right)
	\label{equation:newton-translation}
\label{Fp}
\end{equation}

\begin{equation}
	I_{i}\frac{d\vec{\omega_{i}}}{dt} = 
	\sum_{j \neq i}^{Np} \left(\vec{T}_{c,ij} \right) +
	\sum_{k \neq i}^{Nw} \left(\vec{T}_{c,ik} \right)
\label{Tp}
\end{equation}

\noindent where, for each particle $i$, $\vec{u}_{i}$ is the linear velocity, $\vec{\omega}_{i}$ the angular velocity, $m_{i}$ the mass, and $I_{i}$ the moment of inertia. $\vec{F}_{c,ij}$ and $\vec{F}_{c,ik}$ are, respectively, the contact force between particles \textit{i} and  \textit{j} and the contact force between particle \textit{i} and the wall \textit{k}. $\vec{T}_{c,ij}$ and $\vec{T}_{c,ik}$ are the torque at the particle-particle and particle-wall contacts, respectively, $\vec{F}_{f,i}$ is the force that the fluid exerts on the particle, and $\vec{g}$ is the acceleration of gravity.  We do not consider momentum variations caused directly by the fluid in Eq. \ref{Tp} because they are negligible with respect to contacts \cite{Tsuji, Tsuji2, Liu3}. Contact forces between particles $i$ and $j$ are decomposed in normal and tangential components \cite{Kloss2}, and are given by Eqs. \ref{Fcn} and \ref{Fct}, respectively:

\begin{equation}
	\vec{F}_{cn,ij} = \left( -k_{n} \delta_{n,ij}^{3/2} - \eta_{n} \vec{u}_{ij} \cdot \vec{n}_{ij}  \right) \vec{n}_{ij}
\label{Fcn}
\end{equation}

\begin{equation}
	\vec{F}_{ct,ij} = \left( -k_{t} \delta_{t,ij} - \eta_{t}  \vec{u}_{slip,ij} \cdot \vec{t}_{ij}  \right) \vec{t}_{ij}
\label{Fct}
\end{equation}

\noindent where $n$ and $t$ are subscripts representing the normal and tangential directions, $k$ and $\eta$ are the spring and dashpot coefficients, respectively, $\delta$ represents the overlap between the particles,  $\vec{n}_{ij}$ is the vector that links particle centers, $\vec{u}_{ij}$ is the relative velocity, and $ \vec{u}_{slip,ij}$ is the slip velocity at the contact point. $\vec{t}_{ij}$ is the tangential vector, defined as $ \vec{u}_{slip,ij}/| \vec{u}_{slip,ij}|$. The above relations also hold for particle-wall collisions. The contact forces are modeled following the Hertz-Mindlin and Deresiewicz model \cite{di2004comparison}.

The liquid motion is computed in an Eulerian frame by using the mass and momentum equations given by Eqs. \ref{eq_fluid_mass} and \ref{equation:navier-stokes-momentum}, respectively,

\begin{equation}
	\nabla \cdot \vec{u}_{f} = 0 \qquad \textrm{in $\Omega_{f}$}
\label{eq_fluid_mass} 
\end{equation}

\begin{equation}
	{\frac{\partial \vec{u}_{f}}{\partial t}} + \vec{u}_{f} \cdot
	\nabla \vec{u}_{f}  =  -\frac{\nabla P}{\rho_{f}} + \nu_{f} \nabla^{2}
	\vec{u_{f}} \qquad \textrm{in $\Omega_{f}$}
\label{equation:navier-stokes-momentum}
\end{equation}

\noindent with initial conditions given by Eq. \ref{eq_fluid_ic},

\begin{equation}
	\vec{u}_{f}(\vec{x},t=0) = \vec{u}_{0}(\vec{x}) \qquad \textrm{in $\Omega_{f}$}
\label{eq_fluid_ic}
\end{equation}

\noindent boundary conditions by Eq. \ref{eq_fluid_bc},

\begin{equation}
	\vec{u}_{f} = \vec{u}_{\Gamma} \qquad \textrm{on $\Gamma$}
\label{eq_fluid_bc}
\end{equation}

\noindent and the conditions at the solid-fluid interface by Eqs. \ref{equation:couple-1} and \ref{equation:couple-2},

\begin{equation}
	\vec{u}_{f} = \vec{u}_{s}
	\qquad 
	\textrm{on $\Gamma_{s}$}
	\label{equation:couple-1}
\end{equation}

\begin{equation}
	\vec{\vec{\sigma}} \cdot \vec{n} = \vec{t}
	\qquad
	\textrm{on $\Gamma_{s}$}
	\label{equation:couple-2}
\end{equation}

\noindent where $\Omega_{f}$ is the fluid domain, $\Gamma$ is the CFD boundary, $\Gamma_{s}$ represents the interface between the solid and the fluid, $\vec{u}_{f}$ is the fluid velocity, $\vec{u}_{s}$ is the velocity of the solid particle, $\vec{u}_{\Gamma}$ is the boundary condition for the fluid velocity,  $\vec{u}_{0}$ is the initial condition for the fluid velocity, $\vec{\vec{\sigma}}$ is the stress tensor, $\vec{n}$ is a unit vector normal to the solid surface, $\vec{t}$ is the traction vector of the fluid acting on the solid surface, and $\nu_{s}$ is the kinematic viscosity of the fluid.

Equations \ref{equation:couple-1} and \ref{equation:couple-2} are responsible for the coupling between both phases: Eq. \ref{equation:couple-1} matches the velocity between phases (no-slip condition) and Eq. \ref{equation:couple-2} represents the stress that the fluid exerts on the solid particle. Therefore, the force acting on each particle is obtained by integrating the boundary condition given by Eq. \ref{equation:couple-2} over the body's boundary $\Gamma_{s}$, where the velocity used for the force calculation is weighted by the void fraction distribution on the interface. The same weighting average can be applied to the buoyancy force on each particle, when it is considered. Finally, the force contribution caused by the fluid on particle $i$ is given by:

\begin{equation}
	\vec{F}_{f,i} = \sum_{c \in V_{\Omega_S}} \left( - \nabla P + \mu_{f} \nabla^{2} \vec{u}_{f} \right)_{c} V_c
	\label{equation:fluid-force}
\end{equation}

\noindent where $\Omega_{s}$ is the solid domain, $V_{\Omega_S}$ represents all cells covered or partially covered by the solid, $V$ is the cell volume, and the index $c$ stands for ``cell''.

\subsection{Algorithm}

The resolved four-way computations are implemented in CFDEM according with the following algorithm \cite{Kloss2}:

\begin{enumerate}[label=(\roman*)]
	\item DEM outputs the particle positions and velocities at a certain time step, and those values are passed to the CFD code.\label{item:step-1}
	
	\item A void fraction model identifies the regions covered by the particles and their surfaces, and dynamically refines the mesh on those regions.
	
	\item The fluid flow is computed without considering the presence of particles.
	
	\item Particle velocities are corrected in cells where they are present.
	
	\item The force that the fluid exerts on the particle is computed using the velocity and the pressure fields (Eq. \ref{equation:fluid-force}) and passed to DEM.	
	
	\item The flow field is corrected to satisfy mass conservation.
	
	\item The pressure is once more corrected and the routine restarts from step \ref{item:step-1}.
\end{enumerate}

\section{Numerical setup}
\label{sec:setup}

\begin{figure}[ht]
   	\begin{center}
   		\includegraphics[width=0.80\linewidth]{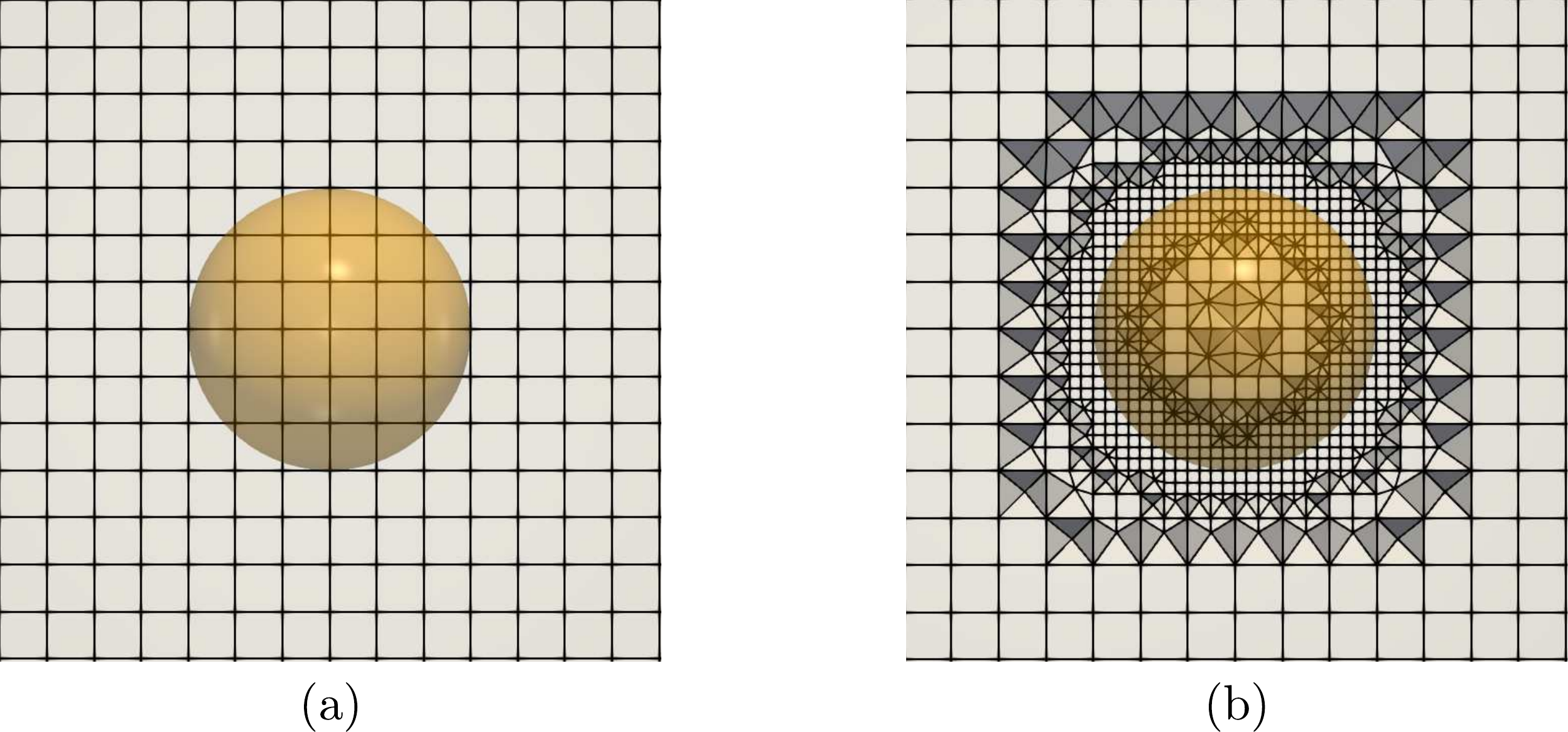}
   	\end{center}
	\caption{(a) Initial refinement; and (b) after the dynamic mesh refinement.}
	\label{figure:mesh-static-dynamic}
 \end{figure}

We make use of the IB method of the open-source code CFDEM \cite{Goniva}, which couples the open-source codes OpenFOAM (www.openfoam.com) and LIGGGHTS \cite{Kloss, Berger}. In the CFD part, the fluid flow is computed using the PISO (pressure-implicit with splitting of operators) algorithm with two main pressure corrections and two non-orthogonal flux corrections. Base meshes consist of hexahedral elements; however, because we use a dynamic grid refinement utility from OpenFOAM that splits some cells, creating tetrahedral elements, the non-orthogonal correctors are necessary. We use linear interpolation (central-difference scheme) to interpolate quantities from the center to the face of cells, and, for the transient term, we use the Euler discretization scheme, which is a first-order, bounded, implicit scheme. Figure \ref{figure:mesh-static-dynamic} shows an example of mesh refinement used in the simulations.

\begin{figure}[ht]
   	\begin{center}
			\includegraphics[width=0.90\linewidth]{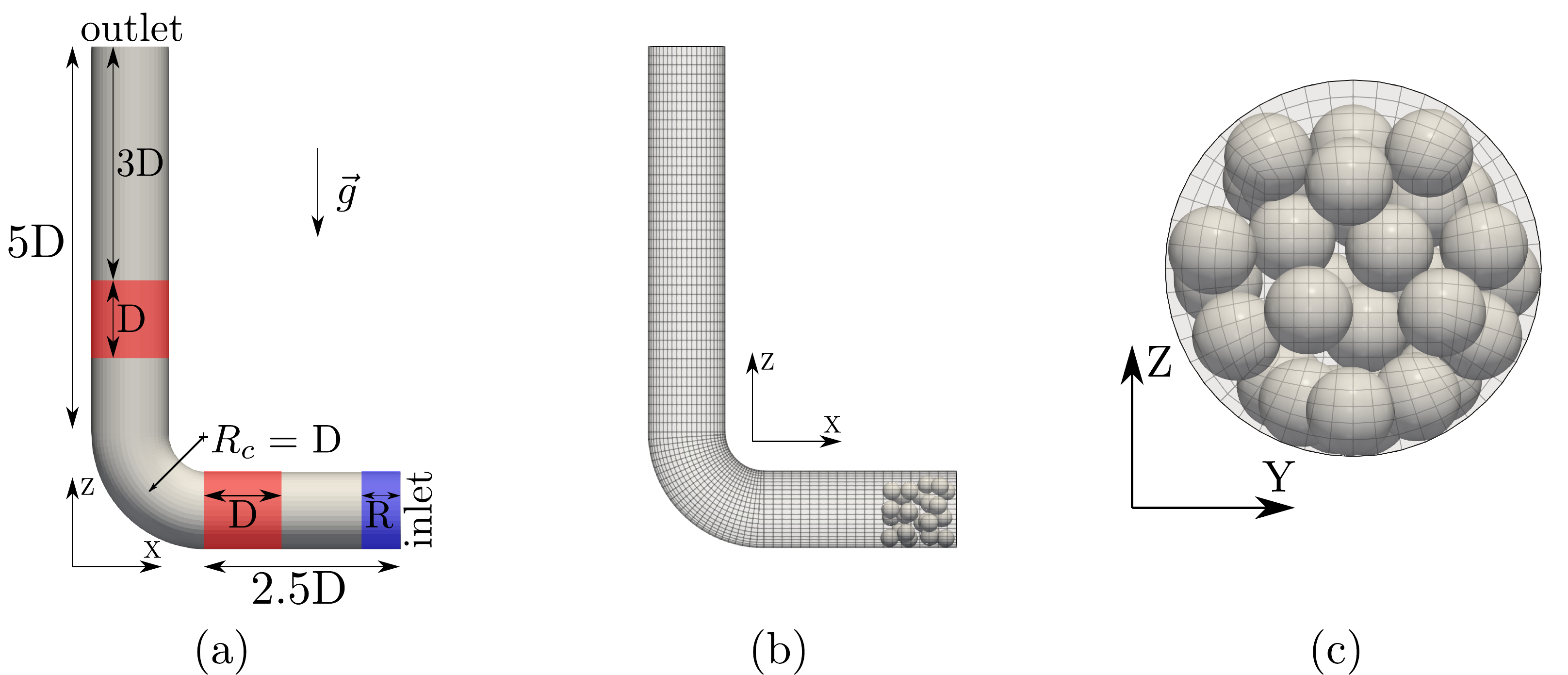}
	\end{center}
	\caption{(a) Dimensions of the CFD domain; (b) side view of the numerical mesh; and (c) front view of the numerical mesh.}
	\label{figure:dimensions-mesh}
 \end{figure}

 Our domain represents a 25.4-mm-ID pipe, with a 2.5$D$-long horizontal section, followed by a 90$^{\circ}$ elbow with a radius of curvature of 1D, and a 5$D$-long vertical section. The inlet is located at the open end of the horizontal section and the outlet at the open end of the vertical section. The CFD domain and initial mesh refinement are shown in Fig. \ref{figure:dimensions-mesh}. The initial mesh refinement is 6 cells per particle diameter in the cross section, but in the flow direction the mesh is coarser. This approach reduces the computational cost and still provides satisfactory results (the validation is presented in Subsection \ref{subsec:validation}). The base mesh consists of 38016 hexahedral elements with a maximum aspect ratio of 7.031815. Initially, there are no particles inside the domain, and when the simulation starts particles with $d$ = 6 mm and $\rho_s$ = 1140 kg/m$^3$ are randomly generated at the inlet, with specific particle rates, for water velocities ranging from 0.06 to 0.10 m/s (0.16 m/s for one specific simulation).

Fluid enters the domain with a fixed value for the velocity and zero-gradient condition for pressure, exits with a zero-gradient condition for the velocity and fixed value for pressure, and has a no-slip condition at the wall. The time steps used for the CFD and DEM were, respectively, 1 $\times$ 10$^{-3}$ and 2 $\times$ 10$^{-5}$ s, leading to a coupling time of 50 DEM time steps. Those time steps were chosen in order to keep the DEM time step less than 10 \% of the Rayleigh time \cite{Derakhshani} and ensure a coupling within 10 and 50 time steps between the DEM and CFD. Particle properties were chosen to represent Nylon 6-6 (our model for organic particles) and are listed in Tab. \ref{table:simulation-parameters}. The DEM time step was small enough to capture the collision span, which is related to the particle Young's modulus \cite{Tsuji, Tsuji2}. For this reason, the Young's modulus used in the simulations is two orders of magnitude lower than the real value, enhancing the DEM time step while still capturing the particle dynamics \cite{Mondal}. An example of simulation is available in Ref. \cite{Supplemental}, containing CFDEM input and output files for a case with particle rate equal to 50 particles/s and fluid velocity of 0.06 m/s (scripts for post-processing the outputs are also available).

\begin{table}[!h]
	\centering
	\caption{Parameters used in the simulations.}
	\begin{tabular}{m{0.7\linewidth} c}
		\hline
		\hline
		Particle diameter $d$ (mm) & 6 \\
		Particle density $\rho_{s}$ (kg/m$^{3}$) & 1140 \\
		Young's Modulus $E$ (MPa) & 33  \\
		Poisson's ratio $\nu$ & 0.41 \\
		Coefficient of restitution $e$ & 0.5 \\
		Friction coefficient $\mu$ & 0.25 \\
		Fluid density $\rho_{f}$ (kg/m$^{3}$) & 1000 \\
		Dynamic viscosity of the fluid $\mu_{f}$ (Pa$\cdot$s) & 0.001 \\
		\hline
	\end{tabular}
\label{table:simulation-parameters}
\end{table}

\subsection{Numerical validation}
\label{subsec:validation}

We validate the numerical setup and verify mesh convergence by reproducing some experimental results found in the literature: (i) the settling of a sphere in a box as measured experimentally by Ten Cate et al. \cite{ten2002particle}; (ii) the drag coefficient of a fixed sphere as measured experimentally by Brown and Lawler\cite {brown2003sphere}; (iii) the sedimentation velocity of a suspension of particles as measured by Richardson and Zaki \cite{Richardson}.

\subsubsection{Terminal velocity of a free-falling sphere}
\label{subsubsec:validation1}

\begin{figure}[ht]
   	\begin{center}
		\includegraphics[width=0.70\linewidth]{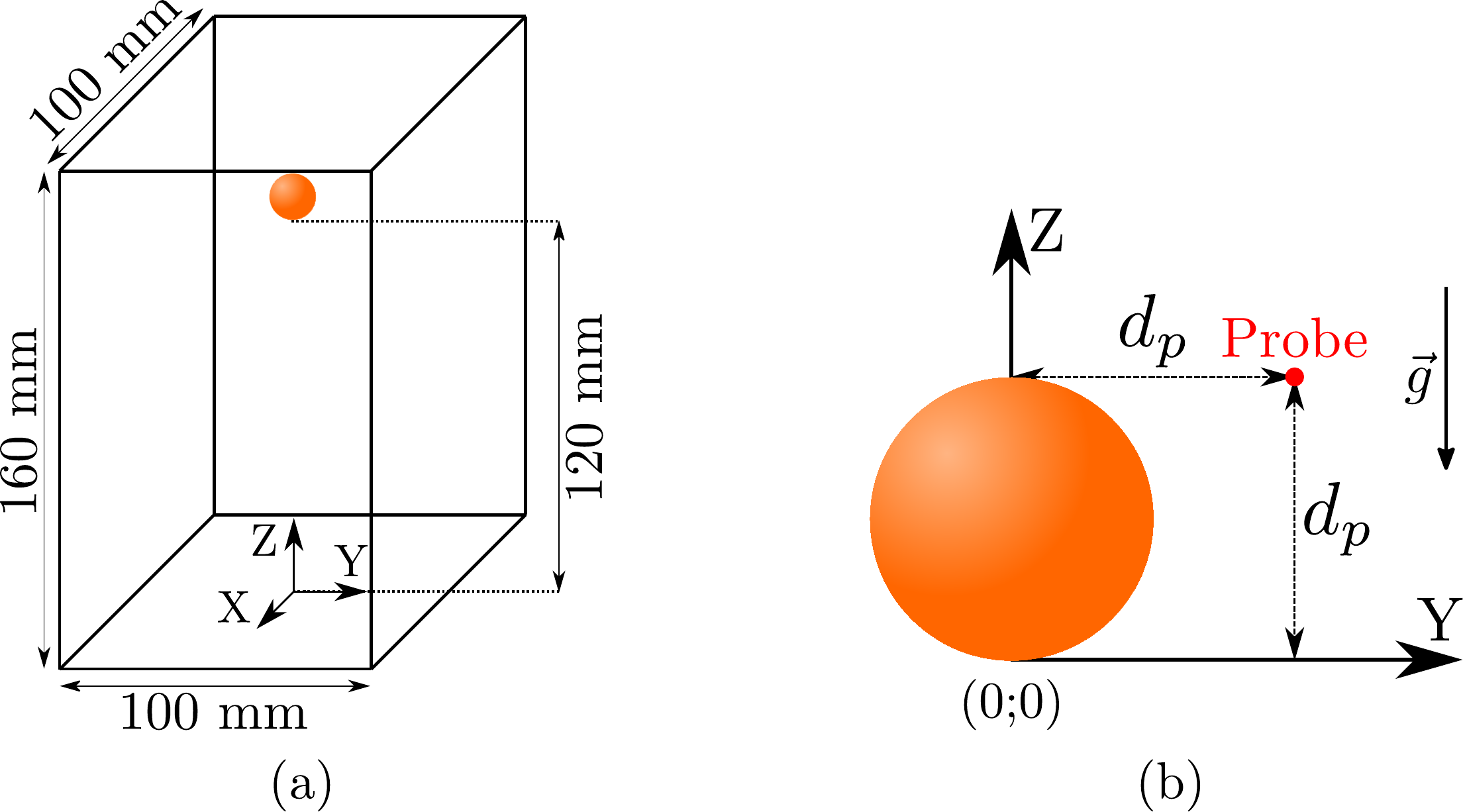}
	\end{center}
\caption{(a) Setup reproducing the experiments. (b) Probe location for the fluid velocity plot (available in the Supplementary Material).}
\label{figure:probe-location}
\end{figure}

We first validate the IB method against the sedimentation experiments of Ten Cate et al. \cite{ten2002particle}. The experiment consisted of an immersed single sphere that settled by free fall in a box. The sphere, with $d$ = 15 mm and $\rho_{s}$ = 1120 kg/m$^{3}$, was confined in a box measuring 100 mm long $\times$ 100 mm wide $\times$ 160 mm high, as shown in Fig. \ref{figure:probe-location}a. Table \ref{table:experiments-parameters} presents the parameters of the tests of Ref. \cite{ten2002particle}, which are also used here.

Because in this particular case we are not interested in the collision dynamics, but rather in that of settling, the DEM time step was set to 1 $\times$10$^{-5}$ s and that of CFD to 1 $\times$10$^{-4}$ s. The domain shown in Fig. \ref{figure:probe-location}a was discretized in three different ways in order to produce either 3, 4.5 or 6 cells per particle diameter, labeled, respectively, $h_3$, $h_2$ and $h_1$, and representing a total of 12800, 43200 and 102400 hexahedral elements each. In addition, we made use of a dynamic mesh refinement on the interface of each particle, as shown in Fig. \ref{figure:mesh-static-dynamic}. Table \ref{table:table-GCI} summarizes the results from three simulations for each case, where $f_1$ corresponds to the particle terminal velocity of the most refined mesh and $f_3$ to that of the coarsest one ($f_2$ being the intermediate mesh), $p$ is the order of convergence, $f_*$ is the extrapolated result following the Richardson Method \cite{roache1998verification}, $GCI_{21}$ and $GCI_{32}$ are, respectively, GCI (grid convergence index) \cite{roache1998verification} for meshes $h_1$-$h_2$ and $h_2$-$h_3$, and R is the ratio that, when close to unity, indicates that the results have converged. The results show that convergence, indeed, is reached by increasing the mesh refinement, and they are quite satisfactory and close to the experimental ones \cite{ten2002particle} for an initial mesh refinement of 6 cells per diameter.

\renewcommand{\arraystretch}{1.1}
\setlength{\tabcolsep}{6.5pt}
\begin{table}[h]
		\caption{Experimental parameters of Ref. \cite{ten2002particle}. Fluid density $\rho_f$, fluid viscosity $\mu_f$, magnitude of the terminal velocity $u_{\infty}$, magnitude of the maximum fluid velocity (normalized) $u_{max}/u_{\infty}$, Reynolds number of the sphere at terminal velocity $Re_{\infty}$, and Stokes number at terminal velocity $St_{\infty}$, for each tested case.}
	\centering
	\begin{tabular}{>{\centering}p{0.07\textwidth} c c c c c c}
		\hline
		\hline
		Case & $\rho_f$ & $\mu_f$ & $u_{\infty}$ & $u_{max}/u_{\infty}$ & $Re_{\infty}$ &  $St_{\infty}$ \\
		& $(kg/m^3)$ & $(Pa \cdot s)$ & $(m/s)$ &  - &  - & - \\
		\hline
		E1 & 970 & 373 & 0.038 & 0.947 & 1.5 & 0.19 \\
		E2 & 965 & 212 & 0.060 & 0.953 & 4.1 & 0.53 \\
		E3 & 962 & 113 & 0.091 & 0.959 & 11.6 & 1.50 \\
		E4 & 960 & 58 & 0.128 & 0.955 & 31.9 & 4.13 \\
		\hline
	\end{tabular}
	\label{table:experiments-parameters}
\end{table}

\begin{table}[h]
	\caption{GCI Method applied to the results of terminal velocity for cases E1 to E4 of Ref. \cite{ten2002particle}. $f_1$ to $f_3$ are the particle terminal velocities with meshes $h_1$ to $h_3$, respectively, $p$ is the order of convergence, $f_*$ is the extrapolated result \cite{roache1998verification} $GCI_{21}$ and $GCI_{32}$ are, respectively, GCI \cite{roache1998verification} for meshes $h_1$-$h_2$ and $h_2$-$h_3$, and R is the convergence ratio.}
	\centering
	\begin{tabular}{>{\centering}p{0.07\textwidth} c c c c c c c c}
		\toprule
		& \multicolumn{3}{c}{Terminal velocity (m/s)} &  &  &  &  &  \\\cline{2-4}
		Case & $f_1$ (m/s) & $f_2$ (m/s) & $f_3$ (m/s) & $p$ &  $f_{*} (m/s)$ & $GCI_{21}$ & $GCI_{32}$ & R\\
		\hline
		E1 & -0.0341 & -0.0341 & -0.0336 & 9.00 & -0.0341 & 0.00\% & 0.04\% & 0.906 \\
		E2 & -0.0564 & -0.0561 & -0.0549 & 3.06 & -0.0566 & 0.45\% & 1.10\% & 1.005 \\
		E3 & -0.0854 & -0.0847 & -0.0825 & 1.97 & -0.0864 & 1.46\% & 2.59\% & 1.009 \\
		E4 & -0.1187 & -0.1174 & -0.1145 & 1.31 & -0.1216 & 3.04\% & 4.48\% & 1.011 \\
		\toprule
	\end{tabular}
	\label{table:table-GCI}
\end{table}

For the tested cases, we measured the particle trajectories and the flow field around the spheres, and the agreement with the experiments of Ten Cate et al. \cite{ten2002particle} is good. Figures showing the particle trajectories, the flow field around the particle, and the temporal evolution of the fluid velocity at a probed location are available in the Supplementary Material.

\subsubsection{Drag coefficient of a fixed sphere}
\label{subsubsec:validation2}

\begin{figure}[!h]
	\centering
		\includegraphics[width=0.55\textwidth]{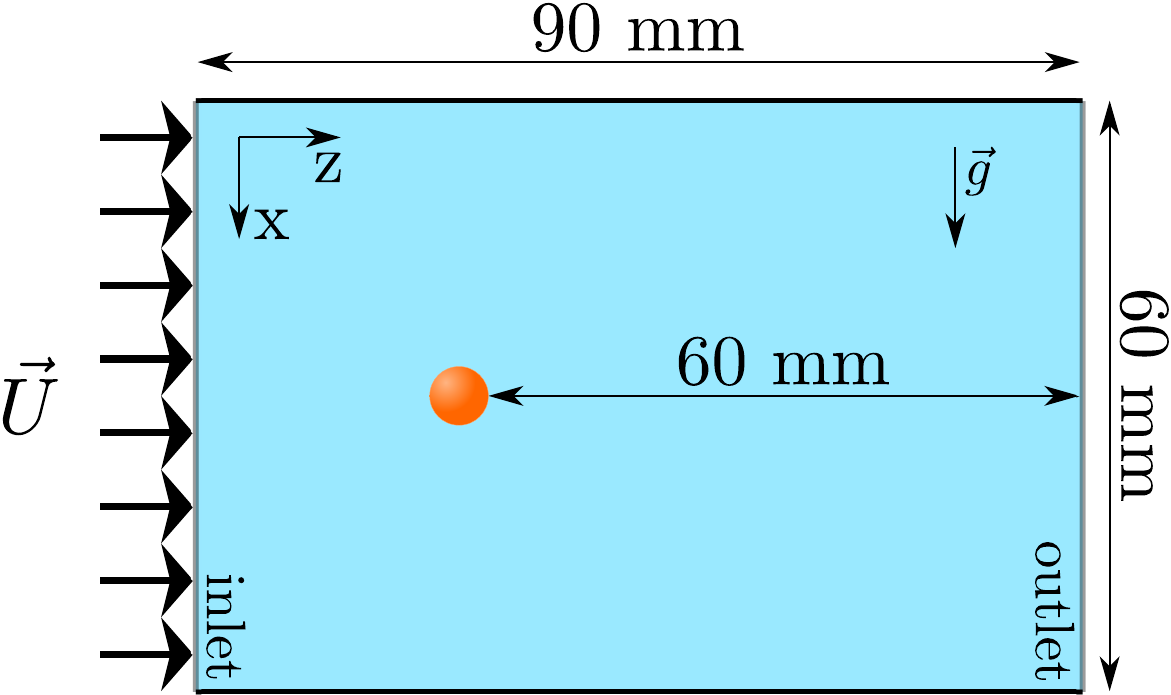}
	\caption{Geometry used for the analysis based on drag coefficient.}
	\label{figure:drag-geometry}
\end{figure}

We next validate the IB computations against values of drag force for a fixed sphere in a fluid flow. The results were obtained for different Reynolds numbers and the comparison with analytical solutions was performed by means of the drag coefficient $C_d$ given by Eq. \ref{eq80}. For that, a particle with $d$ = 6 mm was fixed in a position 60 mm from the outlet, and a horizontal uniform flow was imposed. The fluid density was set to $\rho_f$ = 1000 kg/m$^3$ and its velocity was varied to cover 1 $\leq$ $Re_s$ $\leq$ 1000, where $Re_s$ = $U d / \nu_f$ is the Reynolds number of the sphere, $U$ being the modulus of the free-stream velocity. The geometry and dimensions are depicted in figure \ref{figure:drag-geometry}.

\begin{equation}
	C_{d} = \frac{8 F_{d}}{ \pi d_{p}^{2} \rho_{f} U^{2}}
\label{eq80}
\end{equation}

\begin{figure}[!h]
	\centering
		\includegraphics[width=0.70\textwidth]{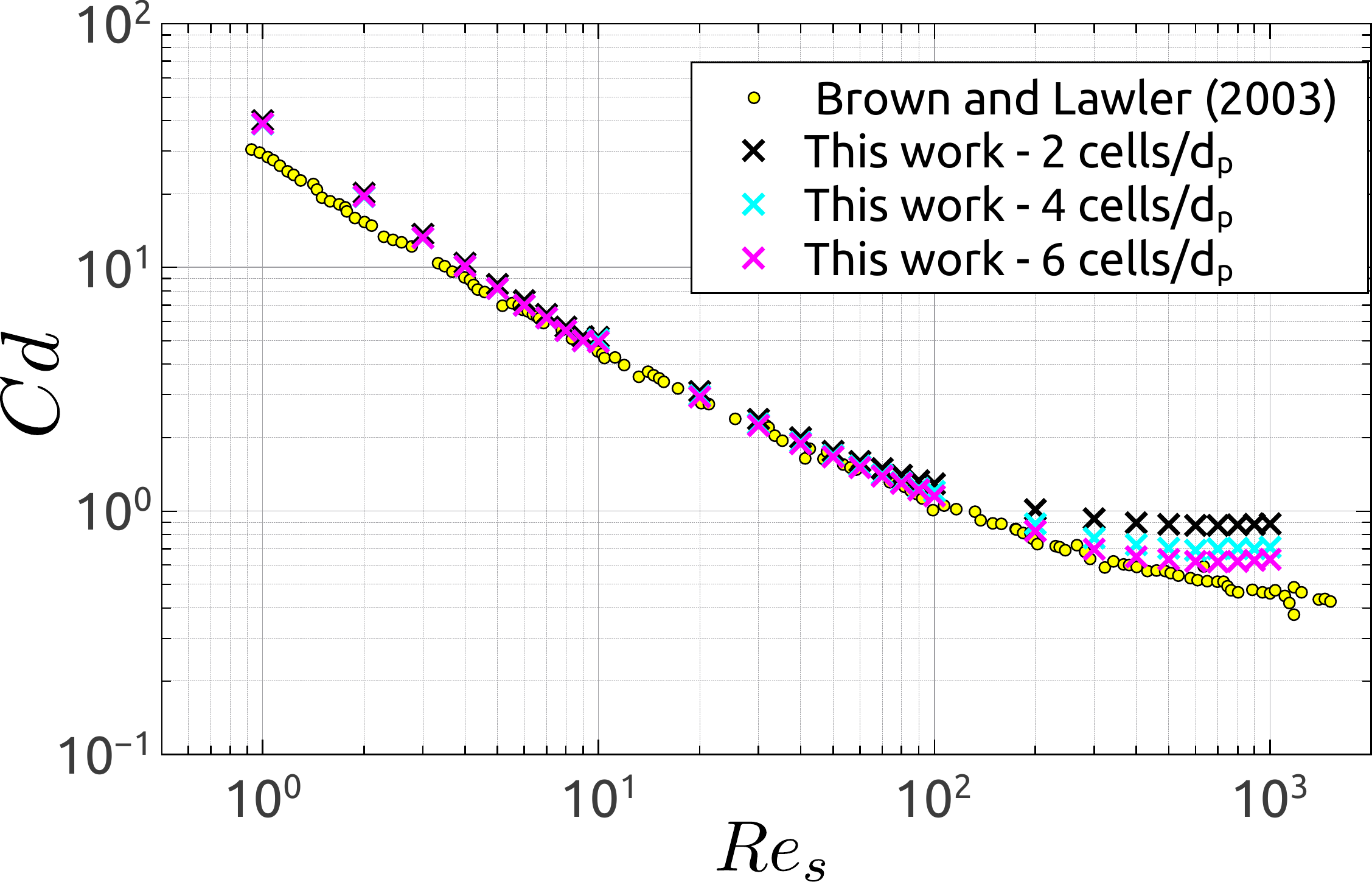}
	\caption{Drag coefficient $C_d$ as a function of the Reynolds number $Re_s$.}
	\label{figure:drag-coeficient}
\end{figure}

The simulations were conducted using three different initial mesh refinements, namely 2, 4 and 6 cells per particle diameter. The CFD time step was adjusted to satisfy the condition of Courant number less than one; therefore, the CFD time steps varied between 5 $\times$ 10$^{-3}$ and 5 $\times$ 10$^{-4}$ s, and those of DEM between 5 $\times$ 10$^{-4}$ and 5 $\times$ 10$^{-5}$ s. The computations were carried out until the velocity and pressure residues were stabilized. The results are presented in Fig. \ref{figure:drag-coeficient}, showing good agreement with the experiments of Brown and Lawler \cite{brown2003sphere}, specially for 5 $\leq$ $Re_s$ $\leq$ 700.

\subsubsection{Sedimentation velocity of a suspension of spheres}
\label{subsubsec:validation3}

Finally, we validate our IB computations against the sedimentation velocity of suspensions of particles as measured by Richardson and Zaki \cite{Richardson}. In a 25.4-mm-ID and 10-$D$-long vertical tube, 49 spheres with $d$ = 6 mm and $\rho_s$ = $1140$ kg/$m^3$ are let to fall in slowly ascending water, for different water velocities (imposed in our simulations). The sedimentation velocity $U_s$ is then computed as the final velocity of particles with respect to water. The mesh and time steps used are basically the same as those described in Section \ref{sec:setup}.

\begin{figure}[!h]
	\centering
	\includegraphics[width=0.50\textwidth]{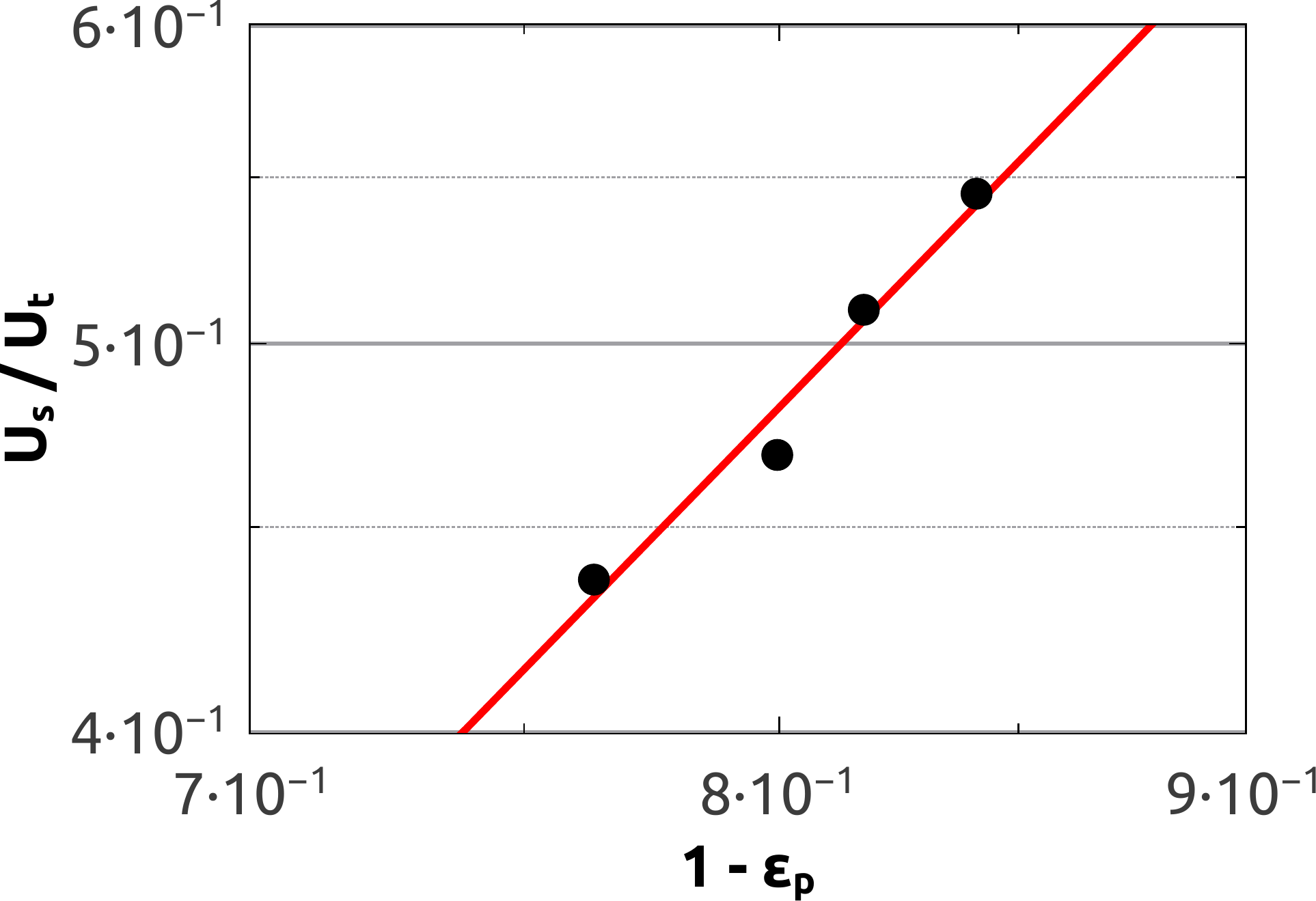}
	\caption{The sedimentation velocity $U_s$ normalized by the terminal velocity $U_t$ as a function of the void fraction (1 - $\varepsilon_p$). The symbols correspond to the numerical outputs and the continuous line to a logarithmic fitting.}
	\label{figure:validation3}
\end{figure}

Figure \ref{figure:validation3} shows the sedimentation velocity $U_s$ normalized by the terminal velocity $U_t$ as a function of the void fraction (1 - $\varepsilon_p$), where $\varepsilon_p$ is the particle fraction. The symbols correspond to the numerical outputs and the continuous line to a logarithmic fitting. In the present case, the terminal velocity $U_t $ = 0.12 m/s was obtained by letting one single particle fall freely in still water, which corresponds to a terminal Reynolds number of $Re_t$ = $U_t d/\nu$ = 720. For this value, the Richardson-Zaki correlation \cite{Richardson},

\begin{equation}
	U_s = U_t \left( 1 - \varepsilon_p \right)^n
	\label{eq:RZ}
\end{equation}
	
\noindent has $n$ = 2.4. From the fitting of our numerical results (Fig. \ref{figure:validation3}), a coefficient $n$ = 2.33 is obtained. By considering that the Richardson-Zaki correlation was obtained empirically and some dispersion has been shown to exist \cite{Difelice, Kramer}, our results agree reasonably well with that correlation.

\section{Results}
\label{section:results}

\begin{figure}[!h]
	\centering
	\includegraphics[width=0.55\textwidth]{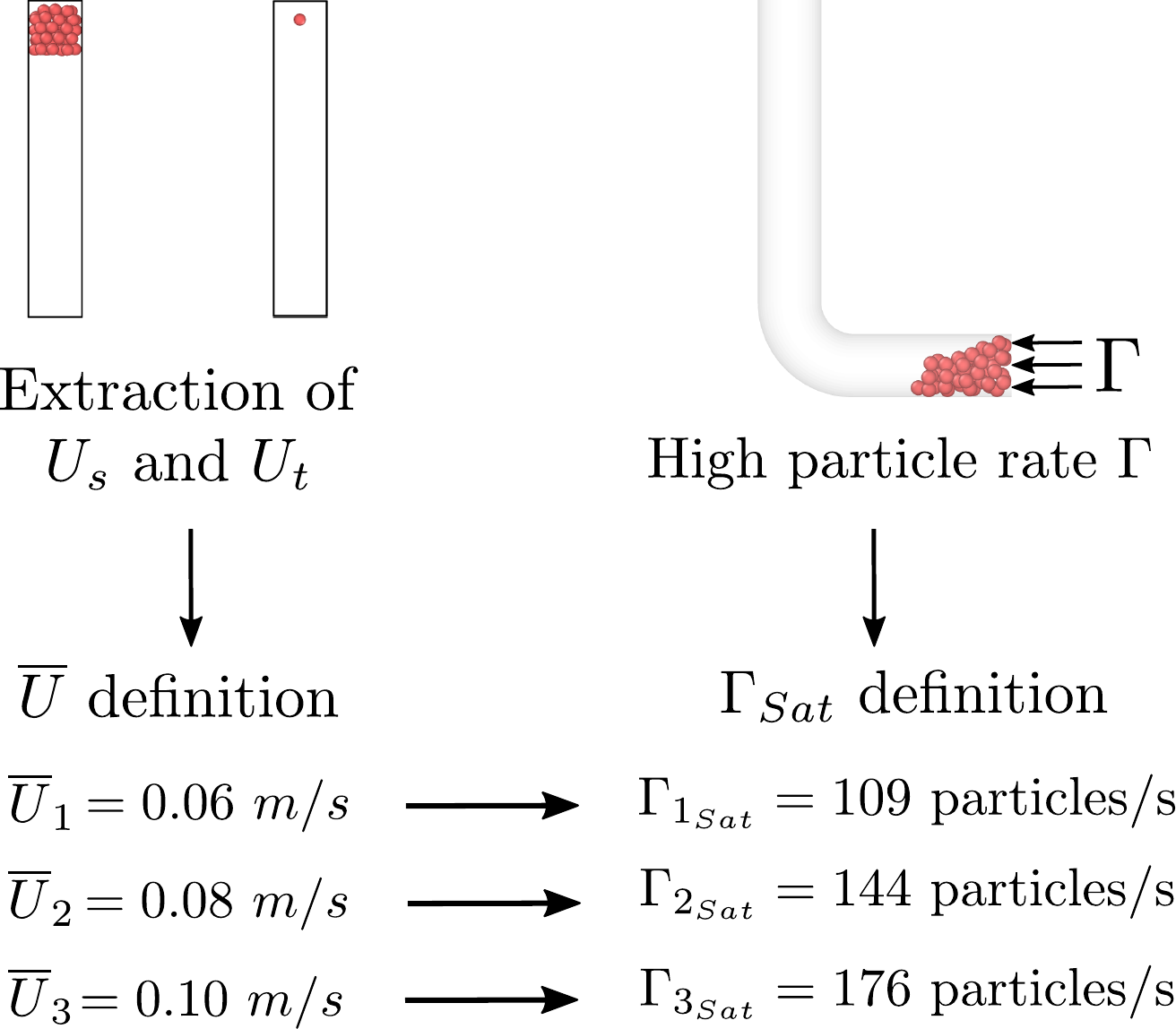}
	\caption{Diagram showing how the water velocities $\overline{U}$ were chosen based on the settling velocity $U_s$, and the maximum particle rate $\Gamma_{max}$ determined.}
	\label{figure:flowchart_proceedure}
\end{figure}

\begin{figure}[!h]
	\centering
		\includegraphics[width=0.50\textwidth]{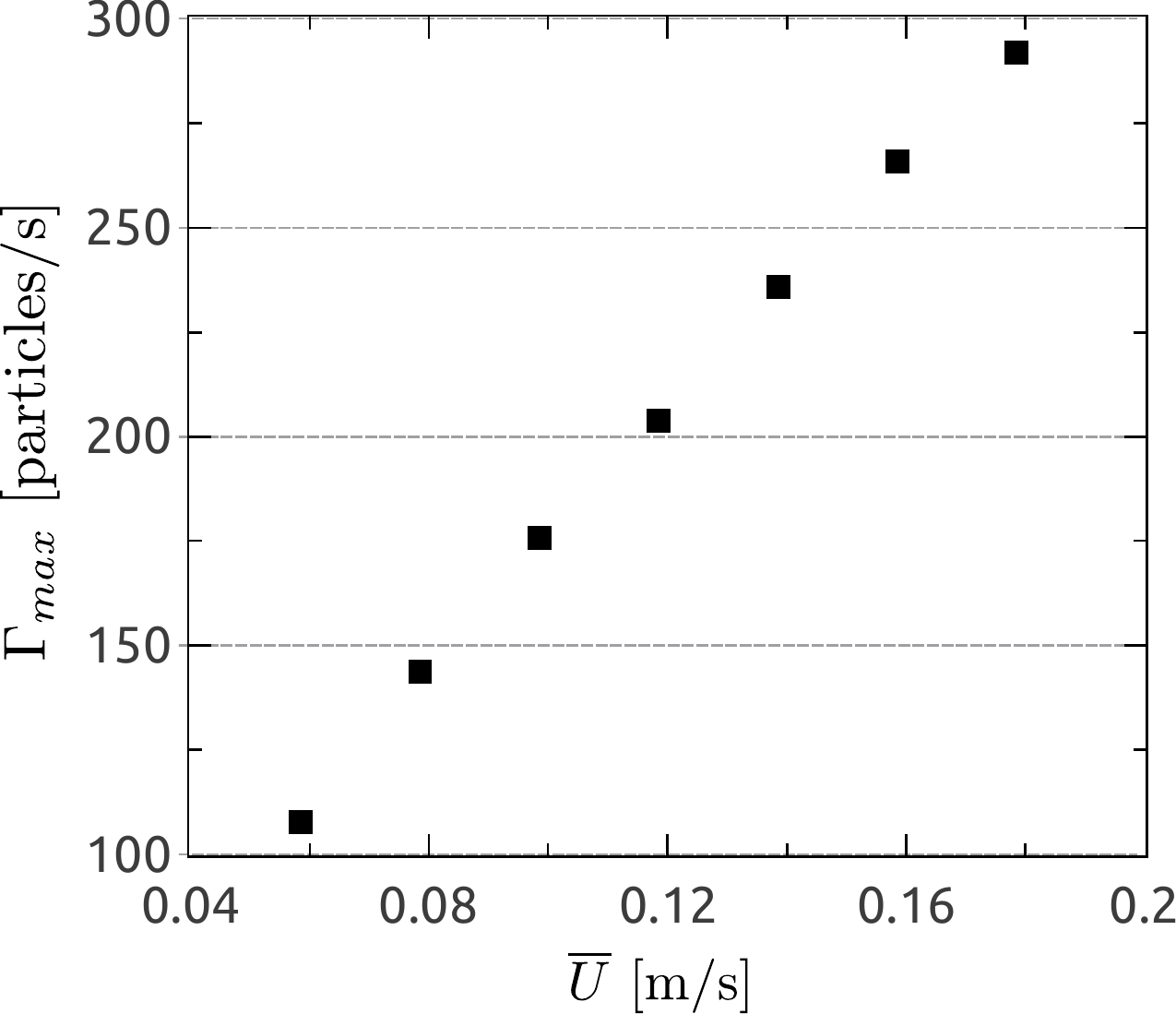}
	\caption{Maximum particle rate $\Gamma_{max}$ as a function of the cross-sectional mean velocity $\overline{U}$.}
	\label{figure:gamma_max}
\end{figure}

We investigate in this paper the effects of the water velocity and particle insertion rate in the transport rate and particle settling. Therefore, before starting the water velocity and insertion rate analyses, we performed two simulations to evaluate the terminal velocity in a 25.4-mm-ID vertical tube of both a single particle and a particle pack with particle fraction of 0.46, and found them to be $U_t$ = 0.12 m/s and $U_s$ = 0.075 m/s, respectively. We then investigate flow conditions in which water velocities are smaller, equal and higher than $U_s$. A very high particle rate was imposed together with the three distinct fluid velocities, and we identified the maximum particle rate that each fluid velocity is able to drag. Once this saturation limit defined, we performed other two simulations for each fluid velocity, with particle rates corresponding to approximately half and a quarter of the respective saturation values. This procedure is illustrated in Fig. \ref{figure:flowchart_proceedure}, which shows how the water velocities were chosen (above and below $U_s$) and the maximum particle rate for each water velocity determined. The parameters of the simulated cases are presented in Tab. \ref{table:data}, where $\overline{U}$ is the cross-sectional mean velocity, $\Gamma$ is the particle rate, and $Re$ = $\overline{U} D/\nu_f$ is the pipe Reynolds number. Clearly, the maximum particle rate increases as the fluid velocity increases, since at higher velocities the fluid can drag more particles. Figures showing both the number of particles inside the tube and the number contacts along time are available in the Supplementary Material, and they show that a steady regime was reached before the end of simulations. In addition to the complete computations analyzed in the following, we performed other four simulations to find the maximum particle rate $\Gamma_{max}$ carried by different fluid velocities. Figure \ref{figure:gamma_max} presents $\Gamma_{max}$ as a function of $\overline{U}$, showing thus the maximum amount of solid particles that can be entrained in the studied pipe for a given flow condition. We observe a roughly linear dependency of $\Gamma_{max}$ on $\overline{U}$, which can be explained by the relatively close values of the densities of particles and water.

\renewcommand{\arraystretch}{1.6}
\setlength{\tabcolsep}{2.5pt}
\begin{table}[!h]
	\centering
	\caption{Parameters investigated for each tested case: cross-sectional mean velocity of the water $\overline{U}$, Reynolds number of the pipe flow $Re$, insertion rate of particles $\Gamma$, flow state, and magnitude of the average velocity of particles.}
	\begin{tabular}{c c c c c c c}
		\toprule
		\multirow{3}{*}{Case} & \multirow{3}{*}{\makecell{$\overline{U}$ \\ (m/s)}} & \multirow{3}{*}{\makecell{$Re$}} & \multirow{3}{*}{\makecell{$\Gamma$ \\ (particles/s)}} & \multirow{3}{*}{\makecell{Flow \\ state}} & \multicolumn{2}{c}{\multirow{2}{*}{\makecell{Magnitude of particle \\ average velocities}}} \\
		& & & & & \\
		& & & & & Horiz. sec.$^{a}$ & Vert. sec.$^{b}$ \\
		\hline
		I & 0.06 & 1524 & 109 & Sat.$^{c}$ & 0.0547 & 0.0526\\
		II & 0.06 & 1524 & 50 & Half-Sat.$^{d}$ & 0.0307 & 0.0440\\
		III & 0.06 & 1524 & 26 & Quarter-Sat.$^{e}$ & 0.0169 & 0.0348\\
		IV & 0.08 & 2032 & 144 & Sat. & 0.0756 & 0.0720\\
		V & 0.08 & 2032 & 66 & Half-Sat. & 0.0436 & 0.0561\\
		VI & 0.08 & 2032 & 37 & Quarter-Sat. & 0.0272 & 0.0457\\
		VII & 0.10 & 2540 & 176 & Sat. & 0.0968 & 0.0920\\
		VIII & 0.10 & 2540 & 79 & Half-Sat. & 0.0623 & 0.0739\\
		IX & 0.10 & 2540 & 46 & Quarter-Sat. & 0.0400 & 0.0556\\
		\toprule
	\end{tabular}
	\label{table:data}
	\\
	\footnotesize{$^a$ Horizontal section; $^b$ Vertical section; $^{c}$ Saturation; $^{d}$ Half-saturation; $^{e}$ One-quarter of saturation;}
\end{table}

The magnitudes of average particle velocities were evaluated at two different regions in the pipe, one in the horizontal and the other in the vertical section, both with a length of 1$D$. The horizontal region is located at a distance between 1.5$D$ and 2.5$D$ from the inlet, and the vertical region at a distance between 3$D$ and 4$D$ from the outlet (red regions in Fig. \ref{figure:dimensions-mesh}a). The averages were computed over all grains crossing those regions during the entire simulation time. We observe that the magnitudes of the particle average velocity in the horizontal section for cases I, IV and VII have values close to those of the fluid at the inlet, and in the vertical section the magnitudes are slightly smaller (around 95\% of those in the horizontal section). Interestingly, when the particle rate is decreased (by half in cases II, V and VIII, and by three-quarters in cases III, VI and IX) the magnitudes in the horizontal section decrease much faster than in the vertical section. Therefore, for $\Gamma$ $<$ $\Gamma_{max}$, average velocities are smaller in the horizontal with respect to the vertical section (approximately 70-80\% for $\Gamma_{max}/2$ and 50-70\% for $\Gamma_{max}/4$).

\begin{figure}[!h]
	\centering
		\includegraphics[width=0.90\textwidth]{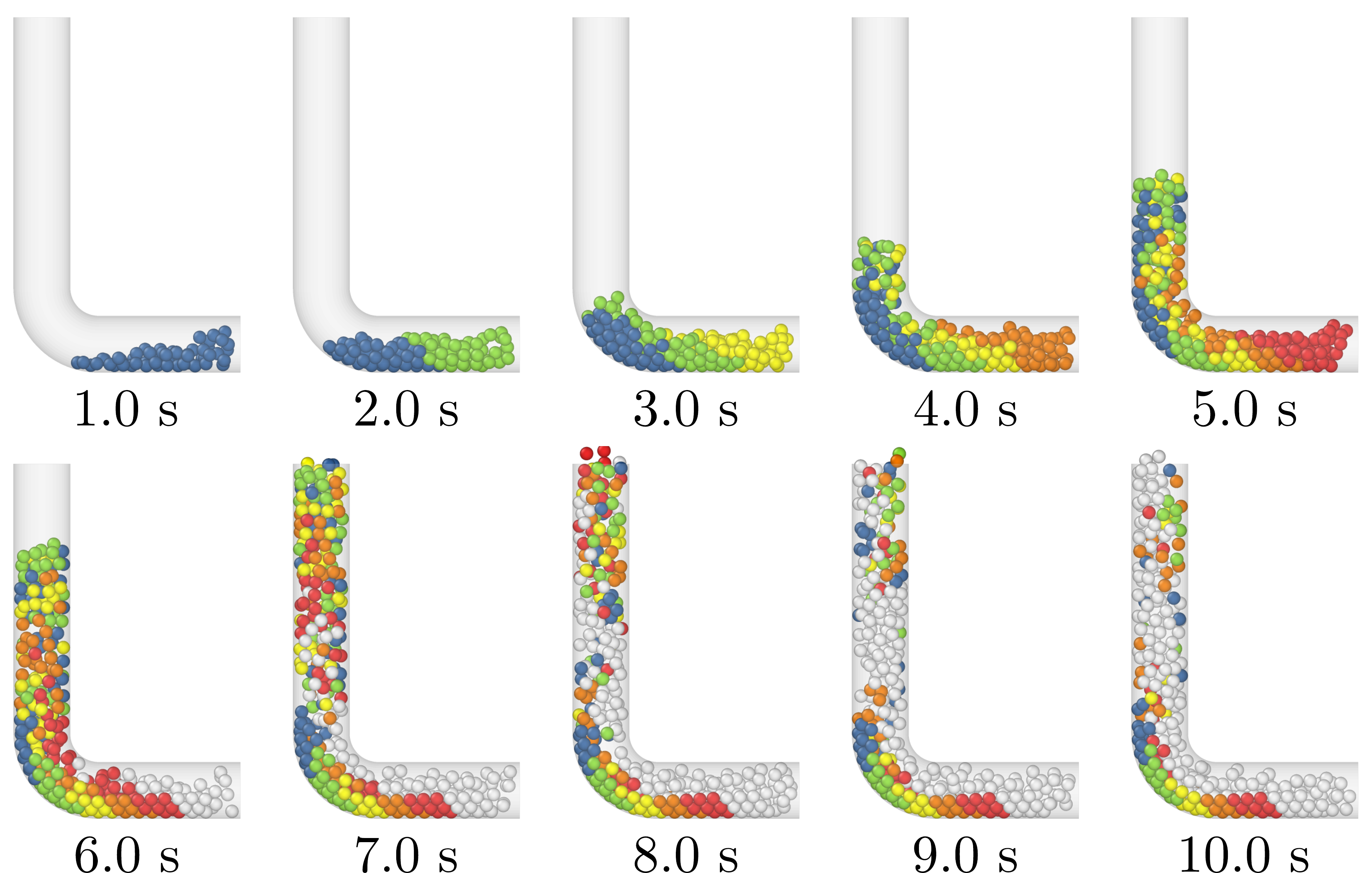}
	\caption{Snapshots of particle positions for Case II (lateral view). Time interval between each frame is 1 s. During 1 s, from $t$ = 0 to 5 s, inserted particles had blue, green, yellow, orange and red colors, respectively. From 5 s on, inserted particles were white.}
	\label{figure:particles-ovito}
\end{figure}

Larger velocities in the vertical than in the horizontal section when the insertion rate decreases indicates that grains are settling in the elbow region. In order to understand that, we followed both the granular structures and individual grains along time. Figure \ref{figure:particles-ovito} presents snapshots of particle positions for Case II for 0 s $\leq$ $t$ $\leq$ 10 s at each 1 s. In the figure, from $t$ = 0 to 5 s, inserted particles had blue, green, yellow, orange and red colors during 1 s, respectively, and from 5 s on the inserted particles were white (see the Supplementary Material for an animation showing the instantaneous positions of particles in the pipe, relative to Fig. \ref{figure:particles-ovito}). We observe that marked particles at the bottom of the horizontal section have low velocities and some of them settle there, while some particles at the top of the horizontal section present higher velocities. At the end of the simulation, particles accumulated on the bottom wall of the elbow consist of groups inserted from 0 to 5 s.

\begin{figure}[!h]
	\begin{center}
			\includegraphics[width=0.80\textwidth]{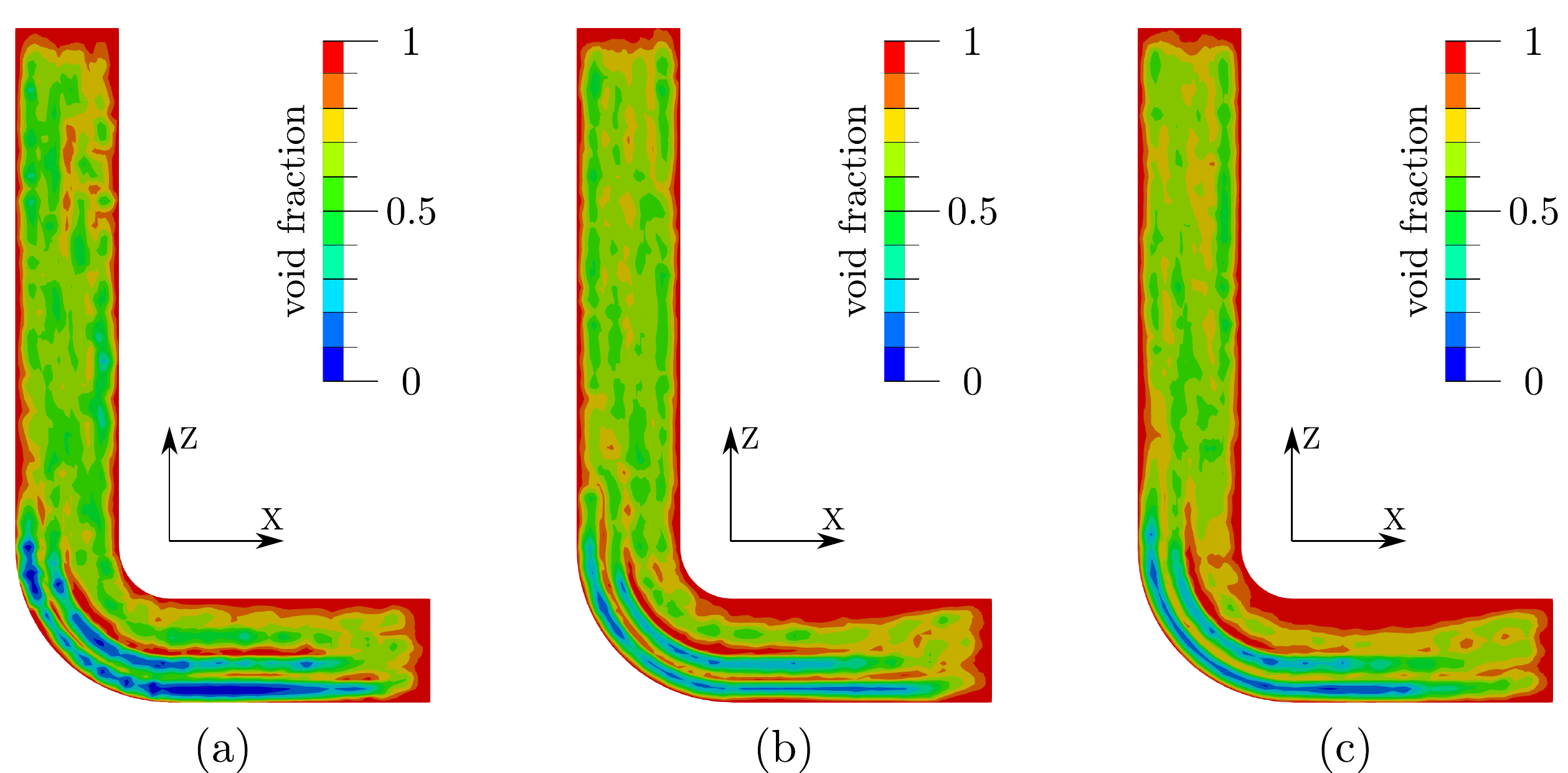}
	\end{center}
		\caption{Time-averaged void fractions in plane $y$ = 0. Figures (a), (b) and (c) correspond to cases II, V and VIII, respectively. Liquid = 1 and solid = 0.}
		\label{figure:sat-voidfraction-lateral}
\end{figure}

\begin{figure}[!h]
	\begin{center}
			\includegraphics[width=0.95\textwidth]{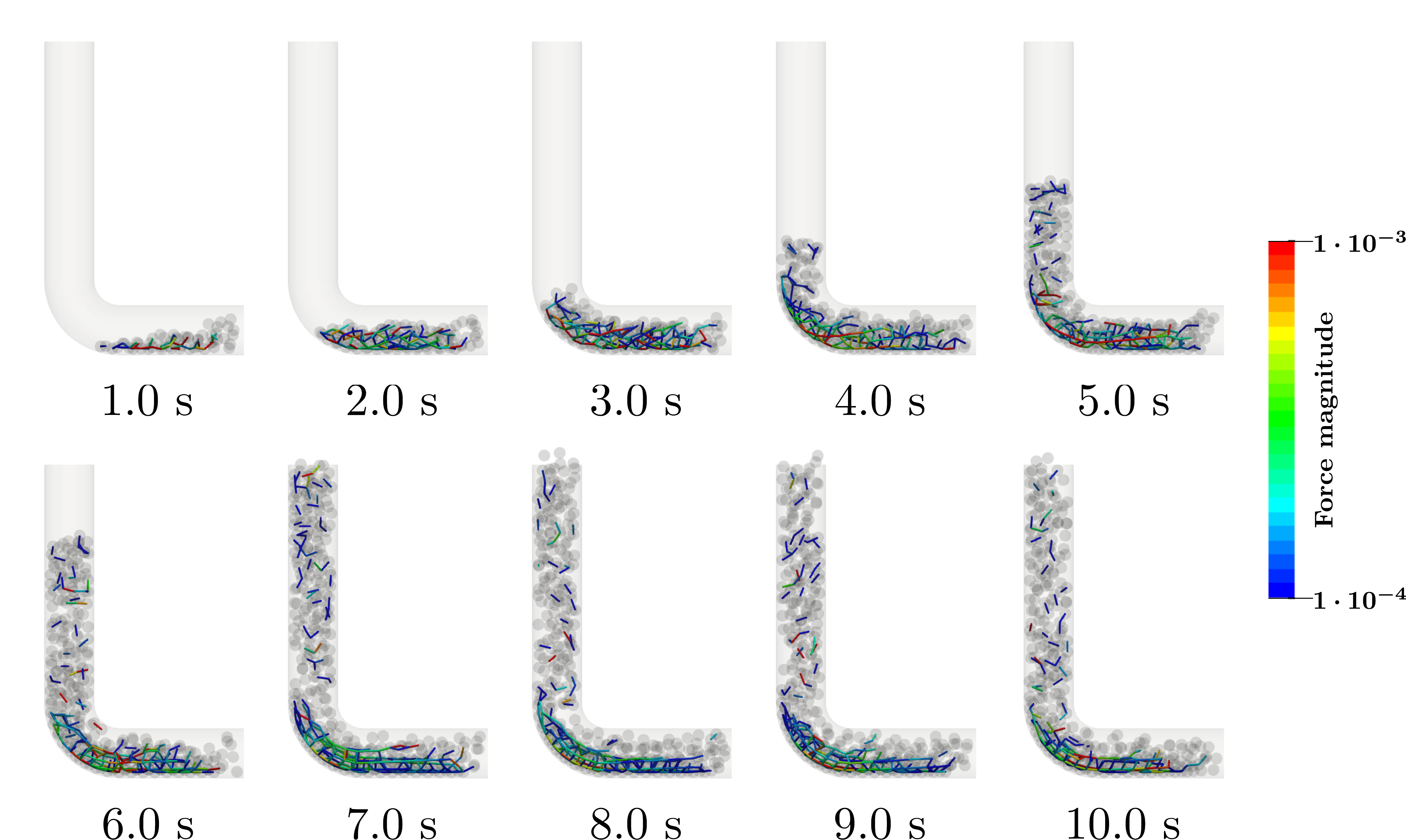}
	\end{center}
	\caption{Snapshots of the network of contact forces for Case II. Time interval between each frame is 1 s and the magnitude is in N.}
	\label{figure:forcechain}
\end{figure}

Figures \ref{figure:sat-voidfraction-lateral}a, \ref{figure:sat-voidfraction-lateral}b and \ref{figure:sat-voidfraction-lateral}c show the time-averaged void fractions (liquid = 1 and solid = 0) in a lateral section of the pipe (plane $y$ = 0) for cases II, V and VIII, respectively. For each case, the average was computed from the instant when the first particle reached the pipe outlet until the end of simulation ($t$ = 10 s). We observe that the low values of void fraction form stripes 1$d$ thick that are aligned at the bottom part of the horizontal section and elbow, indicating that these regions are populated with grains forming an organized structure, and that they become less populated as $\overline{U}$ increases. In addition, we note that four distinct stripes are visible in the vertical section, indicating that grains tend to be distributed radially in that region. In order to inquire into the settling and accumulation of grains, we tracked the network of contact forces within granular structures for Case II and present them in Fig. \ref{figure:forcechain}. We observe the formation of a dense network at the bottom part of the elbow, revealing an organized structure of grains in contact with each other and transmitting the applied stresses within the network. A movie showing the instantaneous positions of grains superposed with the network of contact forces for Case II is available in the Supplementary Material.

\begin{figure}[!h]
	\begin{center}
		\includegraphics[width=0.90\textwidth]{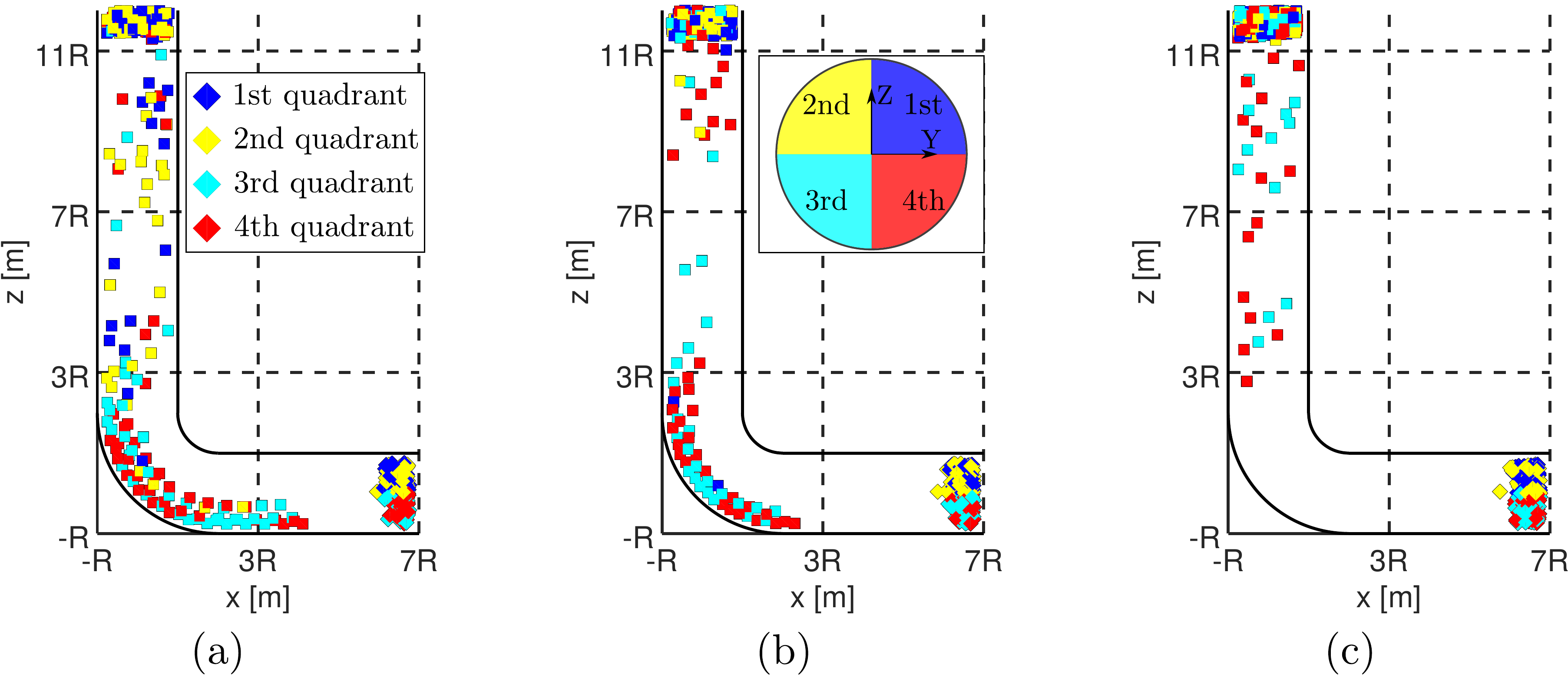}
	\end{center}
	\caption{Initial and final positions of marked particles inserted in $t$ $\leq$ 5 s. Figures (a), (b) and (c) correspond to cases II, V and VIII, respectively.}
	\label{figure:sat-trajectory}
\end{figure}

Figures \ref{figure:sat-trajectory}a, \ref{figure:sat-trajectory}b and \ref{figure:sat-trajectory}c present the initial and final positions of particles for cases II, V and VIII, respectively. The pipe section was divided into four quadrants, as shown in the insert and figure key. In the figures, we show only particles that were inserted in $t$ $\leq$ 5 s, and they are painted in accordance with their positions (quadrant) at the insertion zone, allowing us to investigate the final position of each particle based on its initial location. For the highest fluid velocity, shown in Fig. \ref{figure:sat-trajectory}c, no tracked particle accumulated in the elbow, given the higher particle mobility and lower settling times in this case. Interestingly, in all cases some particles did not reach the end of the vertical section. Instead, they fluctuated in that region rather than going straight to the outlet, in a behavior similar to those of fluidized beds \cite{Cunez, Cunez2}. The number of particles fluctuating in the vertical region decreases as the fluid velocity increases, and they come from different quadrants (see the Supplementary Material for movies showing the behavior of particles).

\begin{figure}[!h]
	\begin{center}
		\includegraphics[width=0.70\textwidth]{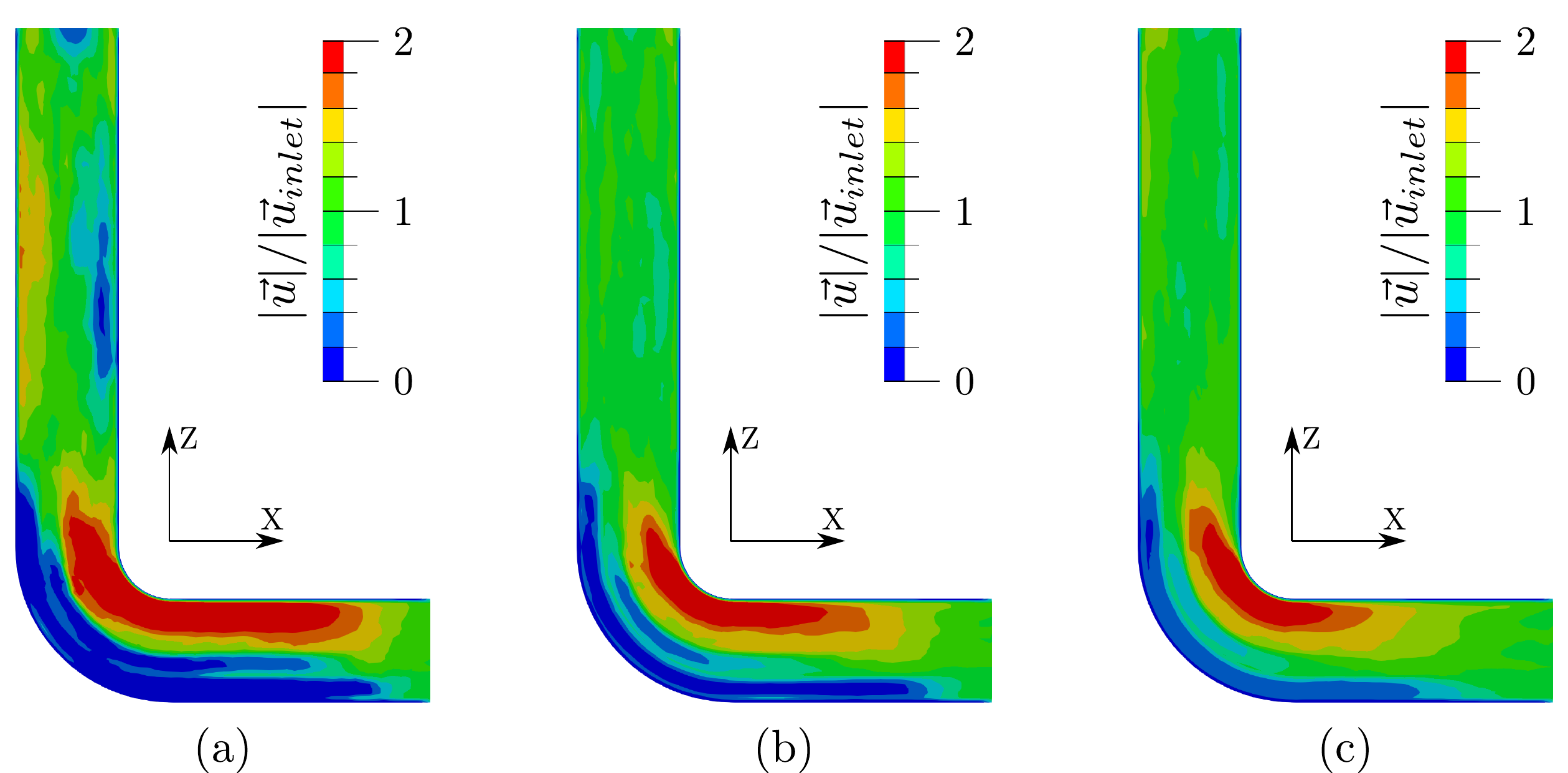}
	\end{center}
		\caption{Time-averaged velocities of the liquid in a lateral section of the pipe (plane $y$ = 0) for cases II, V and VIII, respectively.}
		\label{figure:sat-velocity-lateral}
\end{figure}

\begin{figure}[!h]
	\begin{center}
		\includegraphics[width=0.80\textwidth]{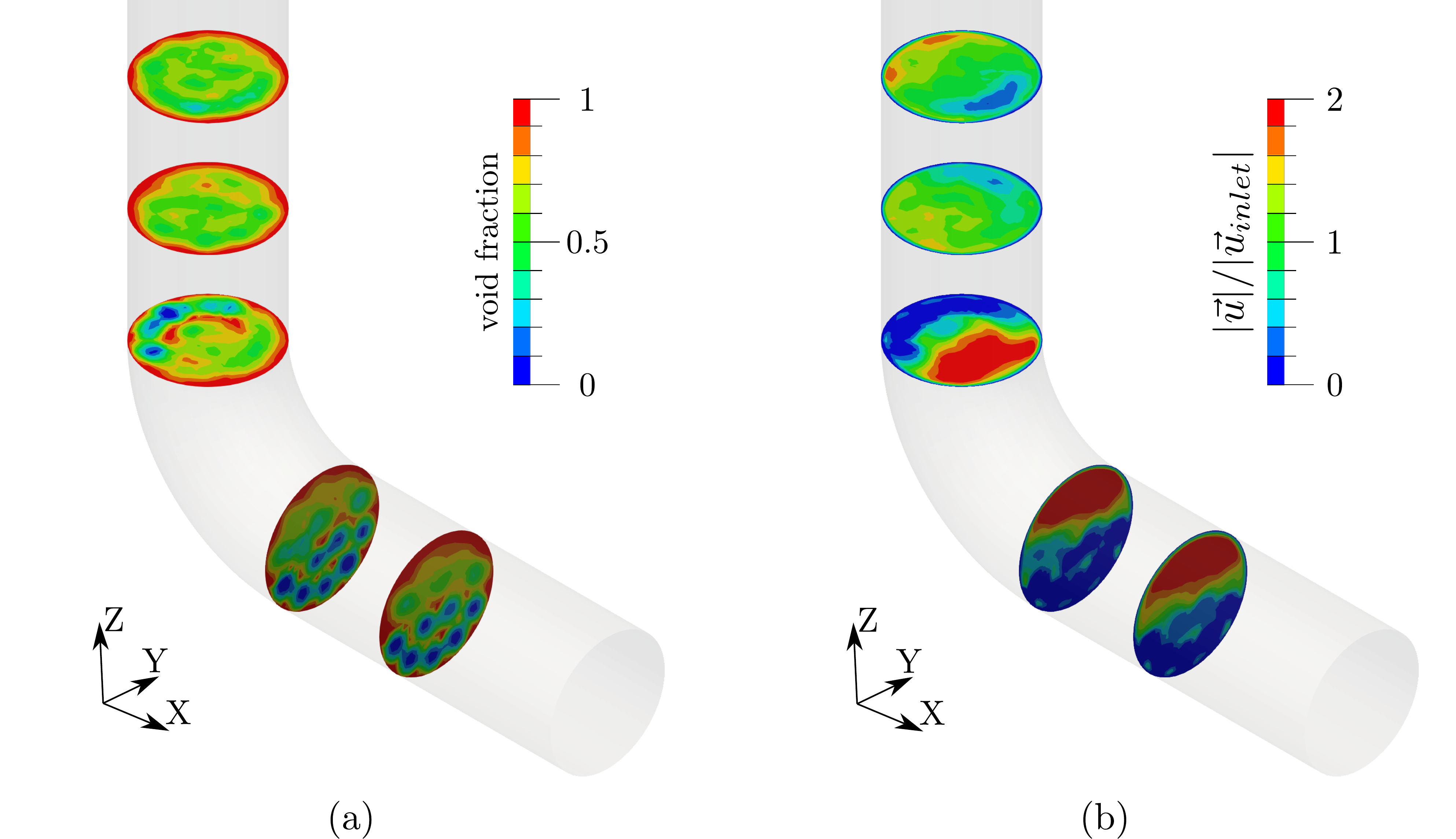}
	\end{center}
		\caption{(a) Time-averaged velocities (magnitude) and (b) void fractions at some pipe cross sections for case II: ($\overline{U}$ = 0.06 m/s).}
		\label{figure:Case-I-perspective}
\end{figure}

We inquire into the liquid flow next. Figures \ref{figure:sat-velocity-lateral}a, \ref{figure:sat-velocity-lateral}b and \ref{figure:sat-velocity-lateral}c show the time-averaged velocities of the liquid in a lateral section of the pipe (plane $y$ = 0) for cases II, V and VIII, respectively, in which the averages were computed over the complete time span. The results corroborate those for the void fraction and particle trajectories, showing an acceleration of the liquid in the upper portion of the horizontal section, and more homogeneous distributions in the vertical section. However, curvature effects caused by the elbow generate an asymmetry in the cross-sections of the vertical portion of the pipe, as can be seen in the cross sections shown in Fig. \ref{figure:Case-I-perspective}. Figures \ref{figure:Case-I-perspective}a and \ref{figure:Case-I-perspective}b show the time-averaged magnitudes of fluid velocities and void fractions at some pipe cross sections for case II. In the vertical portion, besides the curvature effects in fluid velocities, we can notice that particles are organized in radial structures (Fig. \ref{figure:Case-I-perspective}a).

\begin{figure}[!h]
	\centering
		\includegraphics[width=0.80\textwidth]{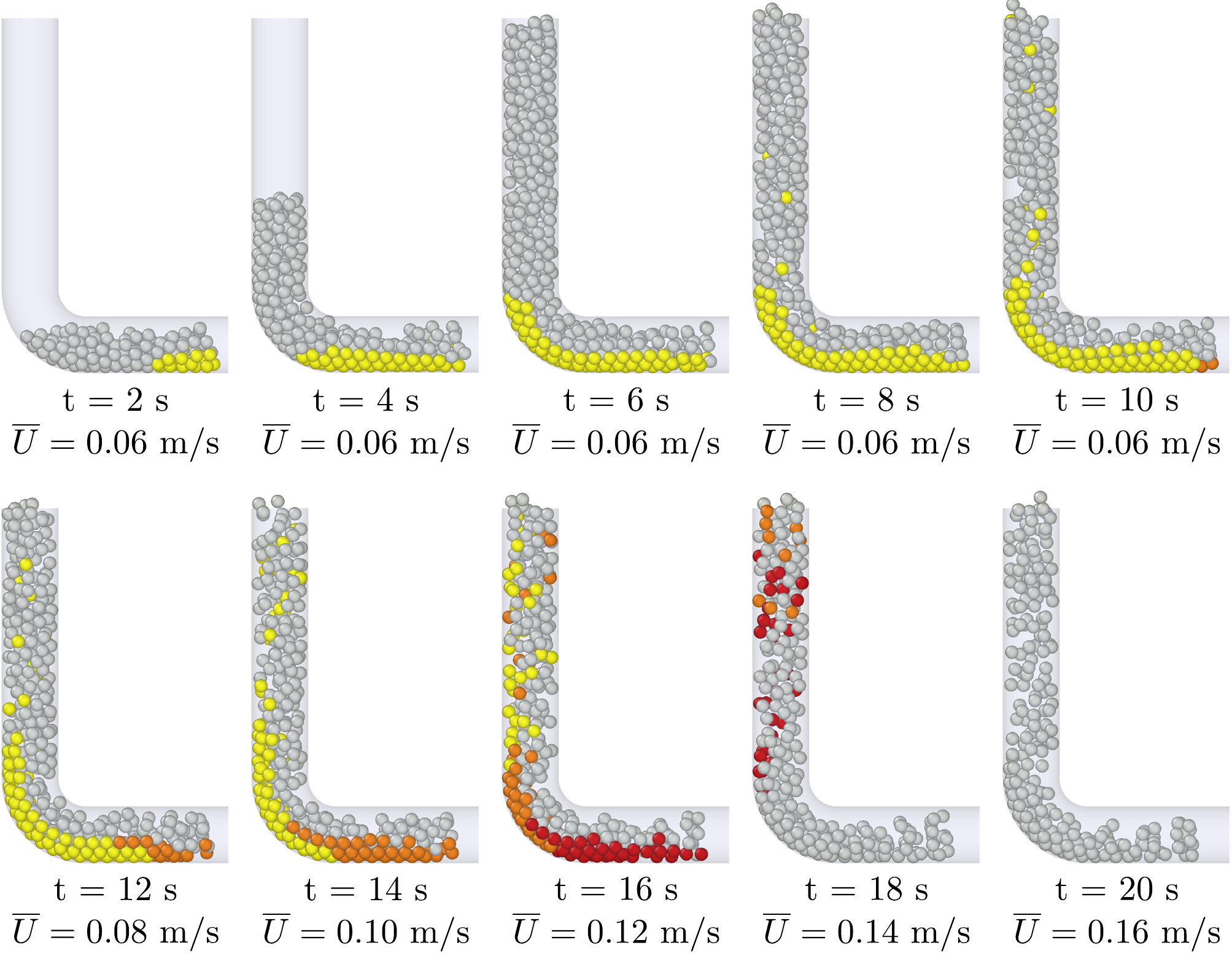}
	\caption{Snapshots of particle positions for increasing $\overline{U}$ from 0.06 m/s (0 s $<$ $t$ $\leq$ 10 s) to 0.16 m/s (10 s $<$ $t$ $\leq$ 20 s). Time interval between each frame is 2 s. Yellow, orange and red particles are those that settled on the bottom of the tube for $\overline{U}$ = 0.06, 0.06-0.10, and 0.10-0.12 m/s, respectively.}
	\label{figure:increasingU}
\end{figure}

Finally, based on the observation that for smaller insertion rates particles settle and tend to accumulate on the bottom of the elbow, we investigate next a simple procedure to remove the settled structure once it has formed. Since decreasing $\Gamma$ worsens the problem, a solution that remains is to impose a transient by increasing $\overline{U}$, even for saturated transport, in order to put settled grains back into motion. In principle, this solution cannot be taken for granted without further investigation because the formation of granular plugs has been reported for heavier particles \cite{Cunez3, Cunez4}. However, the present case concerns much lighter particles that may follow the liquid more closely and not crystallize or jam. Therefore, we simulated Case II (0.06 m/s) for 10 s and, after a settled lattice has formed, increased $\overline{U}$ by steps of 0.01 m/s at 1 s intervals, keeping $\Gamma$ constant. Figure \ref{figure:increasingU} shows snapshots of particle positions for increasing $\overline{U}$ taken at each 2 s, and where yellow, orange and red particles are those that settled on the bottom of the tube for $\overline{U}$ = 0.06, 0.06-0.10, and 0.10-0.12 m/s, respectively. An animation showing the instantaneous positions of particles corresponding to Fig. \ref{figure:increasingU} is available in the Supplementary Material.

From Fig. \ref{figure:increasingU} and the animation in the Supplementary Material, we observe the accumulation of settled particles on the bottom of the elbow when $\overline{U}$ = 0.06 m/s, with the formation of a granular lattice. As $\overline{U}$ increases, particle mobility increases, and at $\overline{U}$ = 0.08 m/s particles from the lattice are gradually entrained by the fluid flow. By increasing further the liquid velocity, larger amounts of particles are dragged by the flow, and from $\overline{U}$ = 0.10 m/s all grains on the bottom of the elbow are moving continuously, i.e., the granular lattice slips over the wall. Finally, when $\overline{U}$ = 0.16 m/s the granular lattice is broken, with grains on the bottom of the elbow presenting higher mobilities and being more spaced, with a consequent decrease in the number of contacts (see the Supplementary Material for a figure showing the number of contacts between grains as a function of time). The formation of granular plugs, crystallization or jamming was not observed. In summary, within the range of fluid, particles and geometry investigated, an increase of roughly 30\% in the liquid velocity is capable of entraining gradually grains from the settled structure, an increase of 70\% keeps that structure slipping over the wall, and an increase of 170\% destroys the structure. Therefore, for piping systems conveying coarse grains of organic matter, this simple procedure can be used to break structures of settled particles in elbows.

\section{Conclusions}

This paper investigated numerically the hydraulic conveying of coarse grains through a very-narrow pipe ($D/d$ = 4.23) curved by a 90$^{\circ}$ elbow that changed the flow direction from horizontal to vertical. The CFD-DEM computations were made for light (organic) materials ($S$ = 1.14), and made use of the IB method of the open-source code CFDEM. We varied the water velocity and particle insertion rate, and identified and tracked the granular structures appearing in the pipe, the motion of individual particles, and the contact network of settled particles. We found the maximum amount of particles transported by a given flow condition, that for either smaller velocities or particle insertion rates more particles settle in the elbow region and form a crystal-like lattice that persists in time, and that particle velocities in the horizontal section decrease much faster than in the vertical section as the particle rate is decreased. We observed also that part of the grains fluctuated in the vertical region rather than going straight to the outlet, in a behavior similar to those of fluidized beds. Finally, once the granular lattice is formed at the bottom of the elbow, we found that increasing the fluid velocity by 30\% causes grains from the settled structure to be entrained by the flow, by 70\% the lattice to slip over the wall, and by 170\% the lattice to be destroyed. Because low insertion rates lead to particle settling, the increase of the liquid velocity can be used to break structures of settled particles in elbows, mitigating the accumulation of coarse grains of organic matter. Our results shed light on how to avoid settling and grain accumulation in elbows and improve the hydraulic conveying of solid particles in industrial facilities.

\section*{Declaration of Competing Interest}

The authors declare no conflict of interest.

\section*{Acknowledgments}

The authors would like to thank the National Council for Scientific and Technological Development - CNPq (Grant No. 144544/2019-9) and the S\~ao Paulo Research Foundation - FAPESP (Grant Nos. 2018/14981-7 and 2019/20888-2) for the financial support provided.



  \bibliographystyle{elsarticle-num} 
  \bibliography{references}






\end{document}